\documentclass[referee]{raa}            
\usepackage{graphicx,times}           
\usepackage{natbib}
\usepackage{amssymb,amsmath}
\usepackage{textcomp}
\usepackage{color}
\usepackage{float}
\bibpunct{(}{)}{;}{a}{}{,}

\usepackage[a4paper=true,pagebackref=true]{hyperref}
\hypersetup{colorlinks = true, linkcolor = green, anchorcolor = red, citecolor = blue, 
filecolor = red, pagecolor = red, urlcolor = red}

\newcommand{\mum}{\ifmmode{\rm \mu m}\else{$\mu$m\ }\fi}

\newcommand{\tabincell}[2]{\begin{tabular}{@{}#1@{}}#2\end{tabular}}

\begin{document}

   \title{X-ray absorption and 9.7 \mum silicate feature as a probe of AGN torus structure}

   \volnopage{Vol.0 (20xx) No.0, 000--000}      
   \setcounter{page}{1}          

   \author{Jun Xu
      \inst{1}
   \and Mouyuan Sun
      \inst{2}
   \and Yongquan Xue
      \inst{1}
   \and Junyao Li
      \inst{1}
   \and Zhicheng He
      \inst{1}
   }

   \institute{Department of Astronomy, CAS Key Laboratory for Research in Galaxies and Cosmology,
             University of Science and Technology of China, Hefei 230026, China; {\it msun88@xmu.edu.cn;xuey@ustc.edu.cn} \\
        \and
       Department of Astronomy, Xiamen University, Xiamen, Fujian 361005, China\\
\vs\no
   {\small Received~~20xx month day; accepted~~20xx~~month day}}

\abstract{
The dusty torus plays a vital role in unifying active galactic nuclei (AGNs). However, the physical structure 
of the torus remains largely unclear. Here we present a systematical investigation of the torus mid-infrared 
(MIR) spectroscopic feature, i.e.,  the 9.7\mum silicate line, of $175$ AGNs selected from the 
\textit{Swift}/BAT Spectroscopic Survey (BASS). Our sample is constructed to ensure that each of the $175$ 
AGNs has \textit{Spizter}/IRS MIR, optical, and X-ray spectroscopic coverage. Therefore, we can simultaneously 
measure the silicate strength, optical emission lines, and X-ray properties (e.g., the column density and the 
intrinsic X-ray luminosity). We show that, consistent with previous works, the silicate strength is 
weakly correlated with the hydrogen column density ($N_\mathrm{H}^\mathrm{X}$), albeit with large scatters. 
For X-ray unobscured AGNs, the silicate-strength-derived $V$-band extinction and the 
broad-H$\alpha$-inferred one are both small; however, for X-ray obscured AGNs, the former is much larger 
than the latter. In addition, we find that the optical type 1 AGNs with strong X-ray 
absorption on average show significant silicate absorption, indicating that their X-ray absorption might not be 
caused by dust-free gas in the broad-line region. Our results suggest that the distribution and structure of the 
obscuring dusty torus are likely to be very complex. We test our results against the smooth and clumpy torus 
models and find evidence in favor of the clumpy torus model. 
}

   \authorrunning{Jun Xu et al.}            
   \titlerunning{Multi-wavelength absorption in BASS AGNs}  

   \maketitle

%
%
\section{Introduction}           
\label{sect:introduction}
The dusty torus, which is widely believed to be responsible for obscuring the broad emission-line region 
and the central engine, is a vital component of the unification models \citep[e.g.,][]{Antonucci1993, 
Urry1995, Netzer2015} of active galactic nuclei (AGNs). However, the structure of the dusty torus remains 
largely undetermined partly because it cannot be readily resolved, despite of a few successful 
attempts \citep[e.g.,][]{Imanishi2018, Garcia-Burillo2019, Gravity2020}. Some 
phenomenological torus models in which the morphology and distribution of dusty clouds are predefined 
are proposed. Then the efforts are focused on solving the sophisticated radiative transfer and obtaining 
the corresponding spectral energy distributions (SEDs) at near- to mid-infrared bands. These models can 
be roughly divided into three categories, i.e., a smooth torus \citep[e.g.,][]{Fritz2006}, a clumpy torus 
\citep[e.g.,][]{Nenkova2008p1, Nenkova2008p2}, or a mixture of these two \citep[e.g.,][]{Siebenmorgen2015}. 

Observationally speaking, the dusty torus manifests itself by various multi wavelength spectroscopic 
signatures. For instance, gas in the dusty torus can induce heavy X-ray obscuration (with X-ray column 
density $N_\mathrm{H}^\mathrm{X}$ to be more than $10^{24}\ \mathrm{cm}^{-2}$) if our line of sight is 
nearly edge-on. The dusty torus can absorb a significant fraction of AGN UV-to-optical continuum emission 
and re-emit mainly at mid-infrared (MIR) bands, thereby making the AGN intrinsic UV-to-optical SEDs much 
redder. In addition to the strong MIR continuum emission, the inner torus region can emit prominent silicate 
emission lines at $9.7\ \mathrm{\mu m}$ and $18\ \mathrm{\mu m}$ due to Si-O stretching and bending 
modes. Such features have indeed been detected by the \textit{Spizter}/IRS spectroscopic observations 
\citep[e.g.,][]{Siebenmorgen2005, Hao2005, Shi2006}. 

According to the simplest AGN unification model, if our line of sight is roughly face-on, we can directly detect 
emission from the central engine, the broad emission-line region, and the inner torus region. Therefore, we 
expect such AGNs to show unambiguous broad emission lines in their UV/optical spectra (classified as optical 
type-1), unobscured X-ray power-law emission and the $9.7\ \mathrm{\mu m}$ and $18\ \mathrm{\mu m}$ silicate 
emission features. In contrast, if the viewing angle is nearly edge-on, our line of sight is obscured by the dusty 
torus. Hence, such AGNs show X-ray spectra with heavy obscuration and lack broad emission lines (classified 
as optical type-2). Their silicate features are also expected to be observed as absorption 
\citep[e.g.,][]{Siebenmorgen2004, Shi2006}. 

In reality, the relations among these dusty torus signatures are much more complex than expected. For example, 
optical type 2 AGNs with silicate emission features have been observed \citep[][]{Sturm2006, Nikutta2009}. 
\cite{Shi2006} systematically investigated the X-ray absorption and the silicate feature of $97$ AGNs 
with various types. They found that there is a connection between $N_\mathrm{H}^\mathrm{X}$ (inferred mostly 
from hardness ratios, which might be biased; see, e.g., \citealt{Li2019}) and the silicate feature (see Eq.~\ref{eq:Ssi} 
for the definition) which is consistent with the expectations of the simplest AGN unification model. However, the 
scatters of the connection are quite large. Therefore, they proposed that the torus structure should be complex and 
clumpy. In addition, \cite{Goulding2012} studied $20$ nearby Compton-thick AGNs and pointed out that, at least for 
Compton-thick AGNs, the observed silicate absorption feature might be caused by galaxy-scale dust rather than a 
compact dusty torus near the central engine. 

The study of dusty torus multi wavelength signatures can benefit from more complete AGN multi wavelength 
surveys. One of such surveys is the \textit{Swift}/BAT all-sky survey \citep{Baumgartner2013}. The resulting AGN 
sample is complete with respect to X-ray absorption since the \textit{Swift}/BAT ultra-hard ($14$--$195$ keV) band 
X-ray can penetrate through obscuring gas clouds with $N_\mathrm{H}^\mathrm{X}>10^{24}\ \mathrm{cm^{-2}}$. 
Thanks to the wide X-ray spectroscopic coverage ($0.3$--$195$ keV), the X-ray properties (especially 
$N_\mathrm{H}^\mathrm{X}$) of each AGN were robustly measured \citep{Ricci2015, Ricci2017b}. Meanwhile, the 
spectroscopic follow-up surveys were performed for a large fraction of this sample \citep[hereafter BASS;\footnote{For 
more details, refer to \url{https://www.bass-survey.com}}][]{Koss2017}. Furthermore, a significant fraction of the 
BASS AGNs also have \textit{Spitzer}/IRS spectroscopic coverage which enables us to examine their silicate features. 
Therefore, this sample is ideal for us to explore the nature and structure of the dusty torus. 

This paper is laid out as follows. In Section~\ref{sect:data}, we describe our sample construction and data analyses. 
In Section~\ref{sect:Results}, we present our results. In Section~\ref{sect:dis}, we discuss the implications of our results. 
Our conclusions are summarized in Section~\ref{sect:sum}.

\section{Sample construction and data reduction}
\label{sect:data}
\begin{figure}
	\centering
	\includegraphics[width=0.75\textwidth,angle=0]{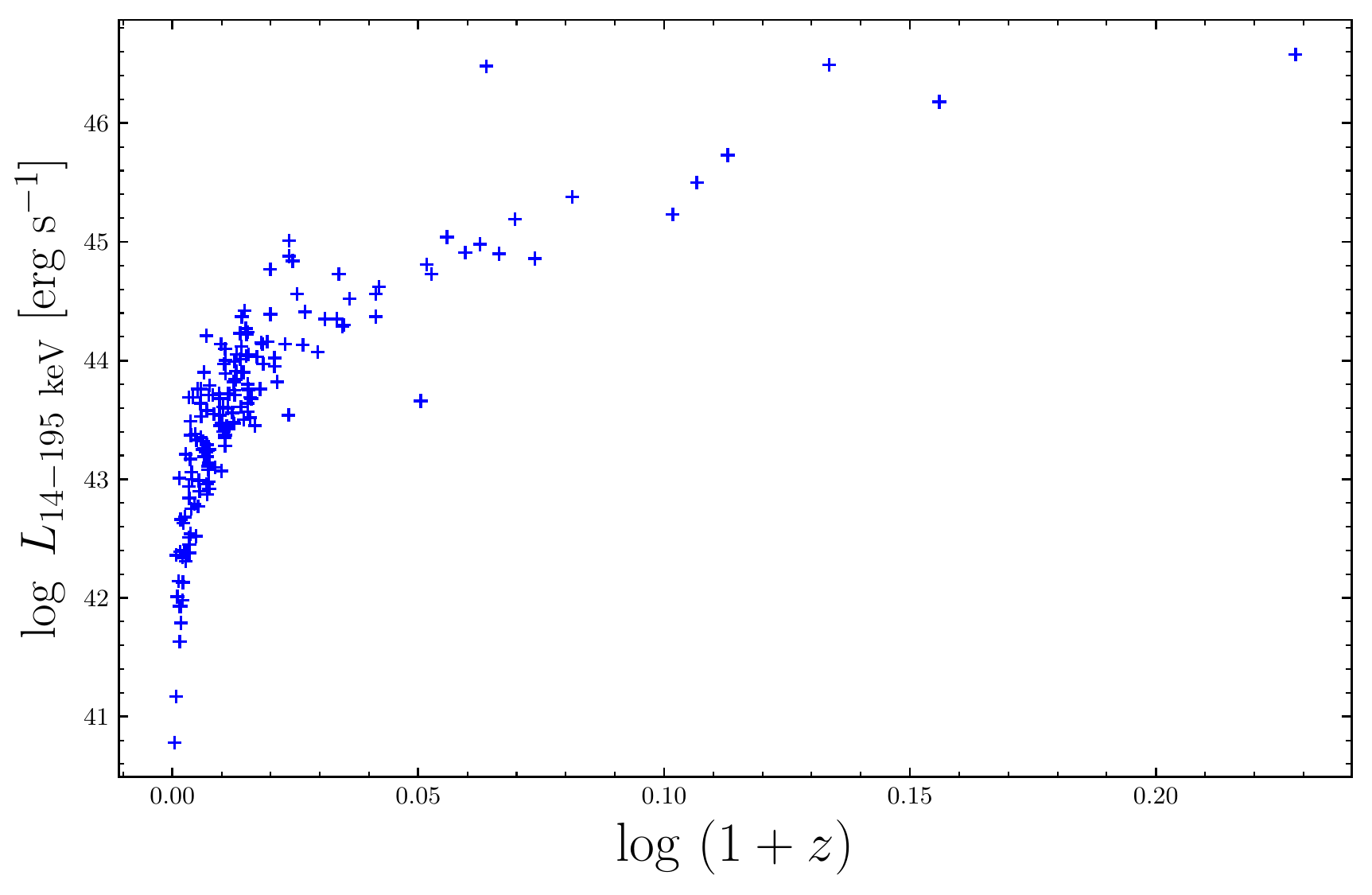}
	\caption{The distribution of our sample in the 14-195 keV X-ray luminosity vs. redshift diagram.}
	\label{fig:Lx_z}
\end{figure}

Our parent sample consists of 836 AGNs from the first $70$-month observations of the unprecedented deep ultra-hard 
X-ray (14--195 keV) survey of the Burst Alert Telescope on the \textit{Swift} Space satellite. Thanks to the wide X-ray 
spectroscopic coverage ($0.3$--$195$ keV), the X-ray properties (e.g., $N_\mathrm{H}^\mathrm{X}$ and the intrinsic 
X-ray luminosity, $L_{14-195\ \mathrm{keV}}$) of all 836 AGNs were well determined and the resulting catalog is publicly 
available \citep{Ricci2015, Ricci2017b}. Therefore, compared with \cite{Shi2006}, our AGNs have more reliable 
$N_\mathrm{H}^\mathrm{X}$ measurements. 

For these 836 AGNs, we use their published counterparts \citep[see][]{Baumgartner2013} to cross-match (by name) 
with the Combined Atlas of Sources with Spitzer IRS Spectra (CASSIS\footnote{For more details, refer to 
\url{https://cassis.sirtf.com/}.}) database \citep{Lebouteiller2011} to construct a new sample, which consists of 208 AGNs. 
Five sources are rejected since their \textit{Spitzer}/IRS spectra do not have spectroscopic coverage around the 9.7 
\mum silicate feature. The remaining 203 AGNs are then cross-matched (by name) with the BASS catalog \citep{Koss2017} 
to obtain their optical spectroscopic measurements. Most of the AGNs ($185/203$) in our sample have optical spectroscopic 
coverage. However, for ten out of the 185 AGNs, their optical measurements and types are absent. Therefore, we reject 
these ten sources. Our final sample, which consists of 175 AGNs, enables us to study the dusty torus in three spectroscopic 
respects, i.e., X-ray absorption, optical type, and the silicate feature. Their luminosity and redshift ranges are presented in 
Figure~\ref{fig:Lx_z}. 

\begin{figure}[H]
	\centering
	\includegraphics[width=0.75\textwidth,angle=0]{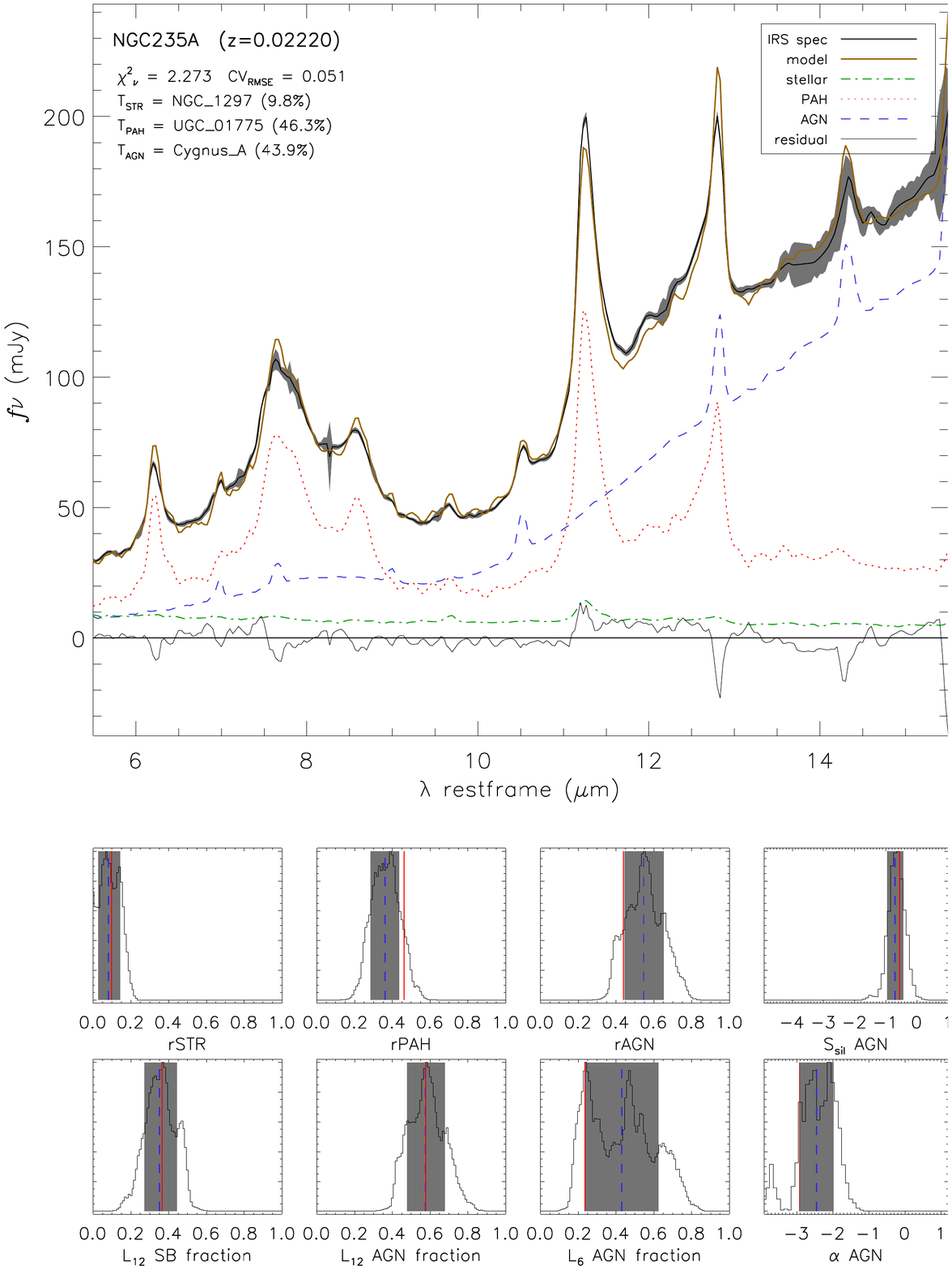}
	\caption{A typical \texttt{deblendIRS} decomposed AGN IRS spectrum (top). The lower panels show the 
		probability distributions of the fitting parameters (for their definitions, refer to Section~\ref{sect:data}), 
		where the red solid and blue dashed lines represent 
		the best-fitting results and the expectations of the distributions. The shaded regions indicate the 1$\sigma$ 
		uncertainties.}
	\label{fig:deblendIRS}
\end{figure}

We perform the $9.7\ \mathrm{\mu m}$ silicate feature measurements by using 
\textit{DeblendIRS}\footnote{For more details, refer to \url{http://www.denebola.org/ahc/deblendIRS/}.} \citep{Hernan2015} 
to fit \textit{Spitzer}/IRS spectra. \textit{DeblendIRS} is an \textit{IDL} package that fits the MIR spectra with a linear combination 
of three spectral templates, i.e., a``pure" AGN template, a ``pure" stellar template, and a ``pure" Polycyclic Aromatic Hydrocarbon 
(PAH, which accounts for the interstellar emission) template. The templates are constructed from real \textit{Spitzer}/IRS spectra 
which are dominated by a single physical component (i.e., AGN, stellar, or interstellar). For each AGN template, the silicate strength 
and the slope ($\alpha$) of a power-law continuum between $8.1$\mum and $12.5$\mum are pre-measured. The silicate strength 
is defined as 
\begin{equation}
S_{\mathrm{Sil}}=\mathrm{ln}\ \frac{F(\lambda_p)}{F_C(\lambda_p)} \\,
\end{equation}
where $F(\lambda_p)$ and $F_C(\lambda_p)$ stand for the maximum flux density of the silicate line profile near 9.7 \mum and the 
correspondingflux density of the underlying continuum profile, respectively. Note that for sources with negative values of 
$S_{\mathrm{Sil}}$, we expect the optical depth of the silicate absorption $\tau_{9.7}=-S_{\mathrm{Sil}}$. Therefore, \textit{DeblendIRS} 
can provide the best-fitting results and uncertainties for the contribution of AGN emission at rest-frame $6$\mum, 12\mum, and 
$5$--$15$\mum (hereafter $L_6$ AGN fraction, $L_{12}$ AGN fraction and rAGN, respectively), the stellar contribution at rest-frame 
$12$\mum and $5$--$15$\mum (hereafter $L_{12}$ SB fraction and rSTR, respectively), the interstellar contribution at rest-frame 
$5$--$15$\mum (hereafter rPAH), $S_{\mathrm{Sil}}$ and $\alpha$ \citep[for more technical details, refer to Section 2 of][]{Hernan2015}.  
An example of our \textit{DeblendIRS} fitting results is shown in the upper panel of Figure~\ref{fig:deblendIRS}. It is evident that the 
best-fitting model explains the data well. 

\begin{figure}[H]
	\centering
	\includegraphics[width=0.75\textwidth,angle=0]{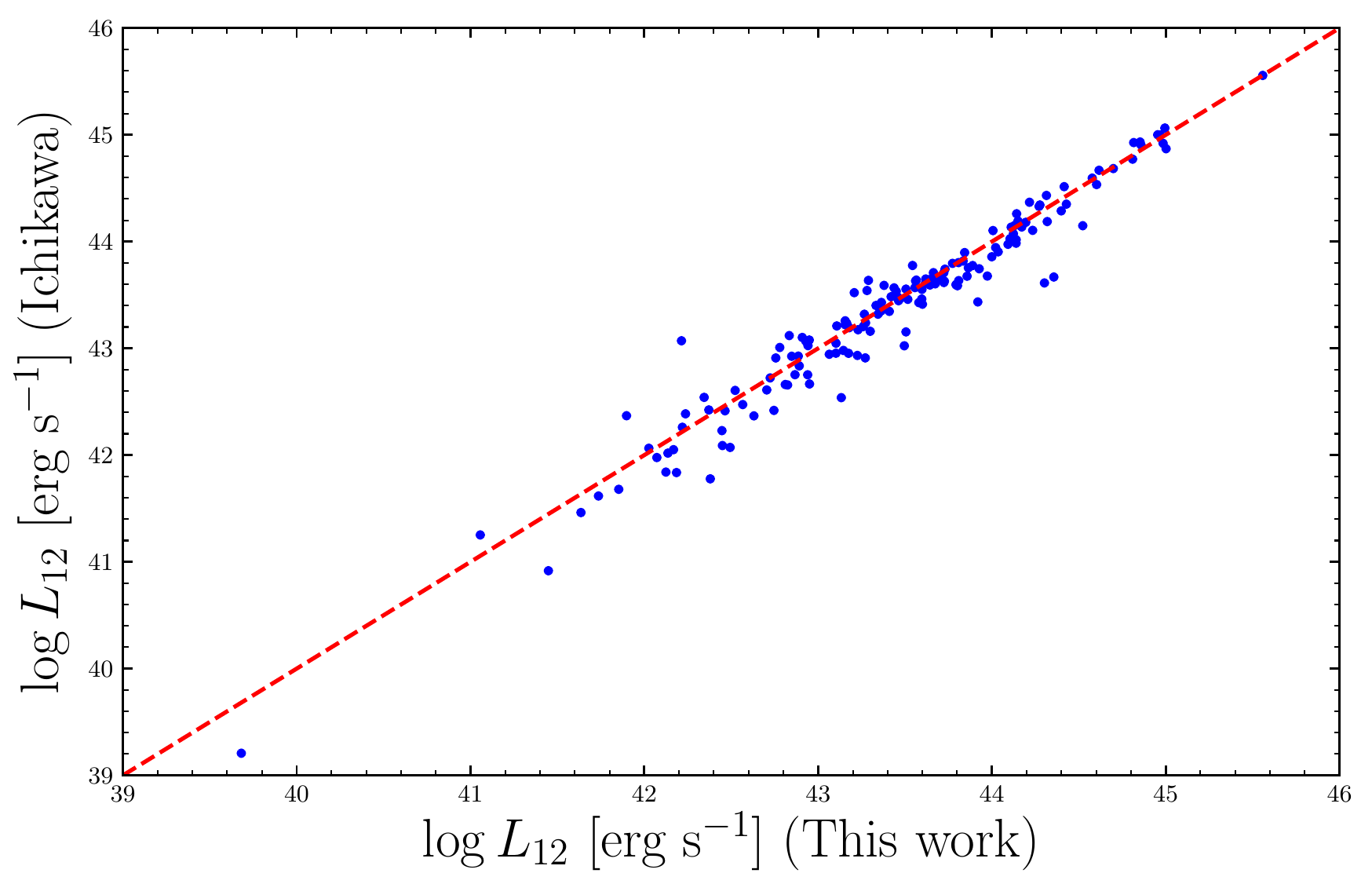}
	\caption{Comparison between our measurements of $\lambda L_\lambda(12\mathrm{\mu m})$ with those of \cite{Ichikawa2019}. 
	The red dashed line indicates the one-to-one relation. Our results are in good aggrement with those of \cite{Ichikawa2019}.}
	\label{fig:12um_comparison}
\end{figure}

Unlike our work, \cite{Ichikawa2019} adopted another \textit{IDL} routine \textit{DecompIR} \citep{Mullaney2011} to decompose the 
multi-band IR (from $\sim 3$\mum to $\sim 200$\mum) photometric data and neglected the \textit{Spitzer}/IRS spectra. There are 
$160$ sources in both our final sample and the catalog of \cite{Ichikawa2019}. To further justify our fitting results, we compare our 
best-fitting 12 $\mu$m monochromic luminosities (hereafter $\lambda L_\lambda(12\mathrm{\mu m})$) with those of \cite{Ichikawa2019} 
(see Figure~\ref{fig:12um_comparison}). Our results are well consistent with those of \cite{Ichikawa2019}; indeed, the median ratio 
between our and their $\lambda L_\lambda(12\mathrm{\mu m})$ is $1.06$. Therefore, we argue that our AGN measurements are 
reliable. 

\section{Results}
\label{sect:Results}
\subsection{Silicate strength and X-ray absorption}
\label{sect:nhx}
\begin{figure}[H]
	\centering
	\includegraphics[width=0.75\textwidth,angle=0]{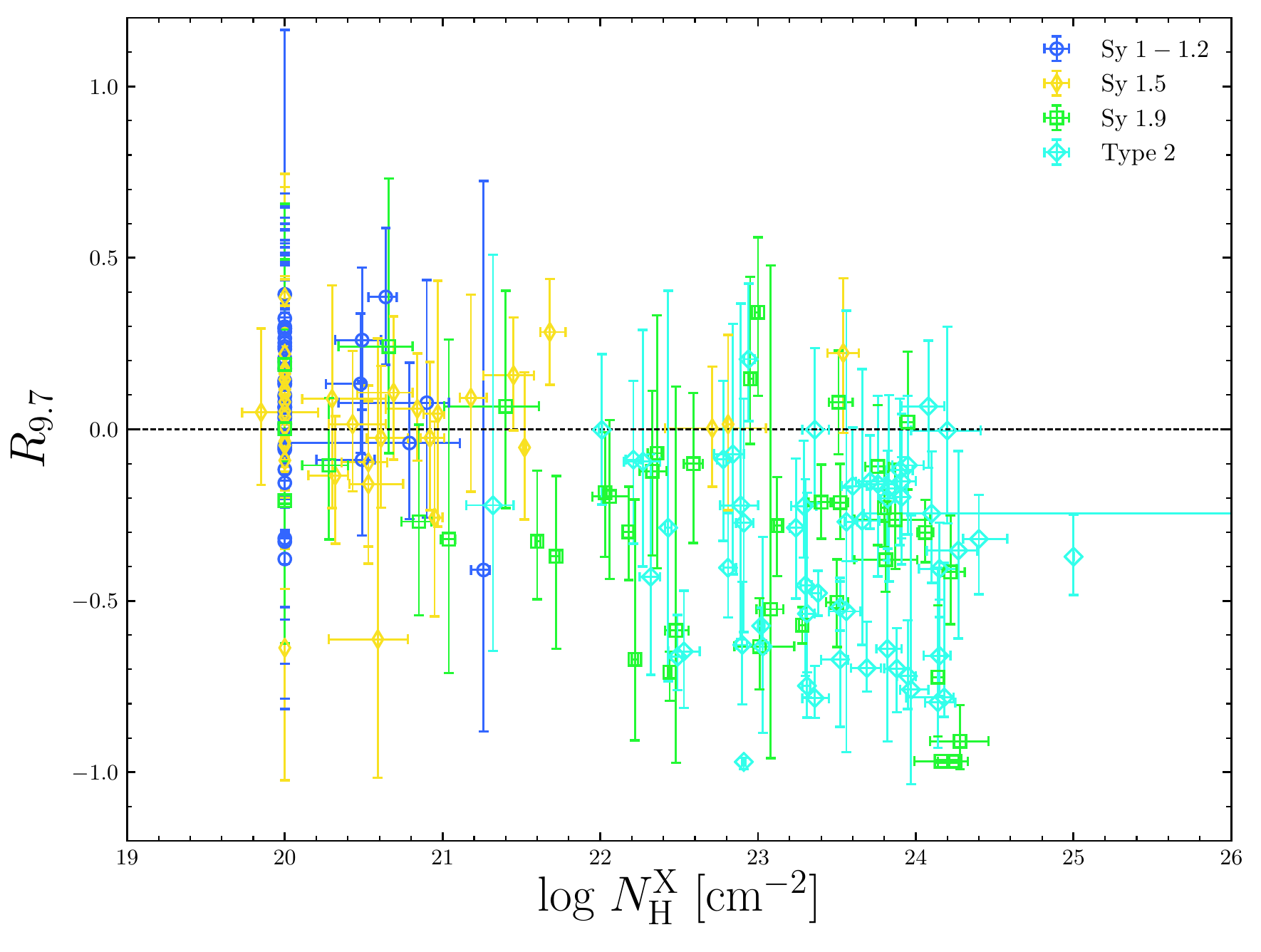}
	\caption{$R_{9.7}$ as a function of $N_{\rm H}^{\rm X}$ for our selected sources. The blue open circles, yellow open diamonds, 
		green open squares, and cyan open diamonds represent type 1-1.2, type 1.5, type 1.9, 
		and type 2 AGNs, respectively.}
	\label{fig:Shi2006fig}
\end{figure}

Following \cite{Shi2006}, we first explore the relationship between the silicate feature and $N_\mathrm{H}^\mathrm{X}$. \cite{Shi2006} 
defined the following quantity, 
\begin{equation}
R_{9.7}=\frac{F(\lambda_p)-F_C(\lambda_p)}{F_C({\lambda_p})}
\label{eq:Ssi}
\end{equation}
where negative/positive $R_{9.7}$ suggests silicate absorption/emission. It is straightforward to show that $R_{9.7}=\exp{(S_{\mathrm{Sil}})} 
- 1$ and approaches $S_{\mathrm{Sil}}$ if $S_{\mathrm{Sil}}$ is close to zero. Figure~\ref{fig:Shi2006fig} plots $R_{9.7}$ as a function of 
$N_\mathrm{H}^\mathrm{X}$ for our final sample. We confirm that, consistent with the result of \cite{Shi2006}, there is a weak anti-correlation 
between $R_{9.7}$ and $N_\mathrm{H}^\mathrm{X}$ (the Spearman's $\rho=-0.63$ and the corresponding $p$-value is $4.5\times 10^{-21}$). 
That is, heavily X-ray obscured AGNs tend to show silicate absorption and absence of broad emission lines, and vice versa. We fit the data 
with the linear relation $R_{9.7}=A + B\log N_\mathrm{H}^\mathrm{X}$ via the MCMC algorithm.\footnote{We use \textit{lnr.py} to perform the 
fit. This \textit{Python} code is available at \url{https://www.astro.princeton.edu/~sifon/pycorner/lnr/}.} The fitting results are $A=4.1^{+0.6}_{-0.5}$ 
and $B=-0.20^{+0.03}_{-0.03}$, which is in agreement with the best-fitting results of \cite{Shi2006}. However, the scatters of the anti-correlation 
are quite large. The large scatters might be caused by several different factors. For example, the gas-to-dust ratio may vary among different 
AGNs. As pointed out by \cite{Shi2006}, the scatter of this ratio should be more than two orders of magnitude, which is unlikely to be the case 
here. Another possibility is that the heavy X-ray absorption is caused by the gas lying closer to the central engine than the dusty torus, e.g., the 
broad emission-line gas. However, this scenario cannot explain the fact that many of these AGNs, which show heavy X-ray obscuration but 
different silicate strengths, are actually optical type-1.9 or type-2 sources, i.e., sources with strong dust extinction. Other possibilities are that the 
dusty torus is not a smooth ``donut'' but a highly clumpy one, the line-of-sight absorption is time-dependent \citep[e.g.,][]{Yang2016, Jaffarian2020}, 
and/or the radiative transfer may also play a role. 

\subsection{Silicate strength and broad-H$\alpha$-line-inferred extinction}
\label{sect:bha}
For a subsample of AGNs with broad H$\alpha$ emission lines, \cite{Shimizu2018} adopted the empirical relation between X-ray and broad 
H$\alpha$ luminosities and the absorption-corrected X-ray luminosities to obtain the intrinsic broad H$\alpha$ luminosities. By comparing 
the intrinsic broad H$\alpha$ luminosities with the observed ones, \cite{Shimizu2018} estimated the optical extinction of the broad-line 
region (hereafter $\mathrm{A_V}[\mathrm{bH}\alpha]$). Then, they explored the relation between $\mathrm{A_V}[\mathrm{bH}\alpha]$ and 
$N_\mathrm{H}^\mathrm{X}$ and found that a significant fraction of AGNs have orders of magnitude higher $N_\mathrm{H}^\mathrm{X}$ 
than the $\mathrm{A_V}[\mathrm{bH}\alpha]$-inferred values by assuming a Galactic ratio of $N_\mathrm{H}$ to $\mathrm{A_V}$ 
\citep{Draine2011}. The population of optical type 1 AGNs with heavy X-ray absorption has also been explored by \cite{Merloni2014} who 
used the $1310$ XMM-COSMOS AGNs as well as several previous works \citep[e.g.,][]{Burtscher2016, Schnorr2016}. 

\begin{figure}[H]
	\centering
	\includegraphics[width=0.75\textwidth,angle=0]{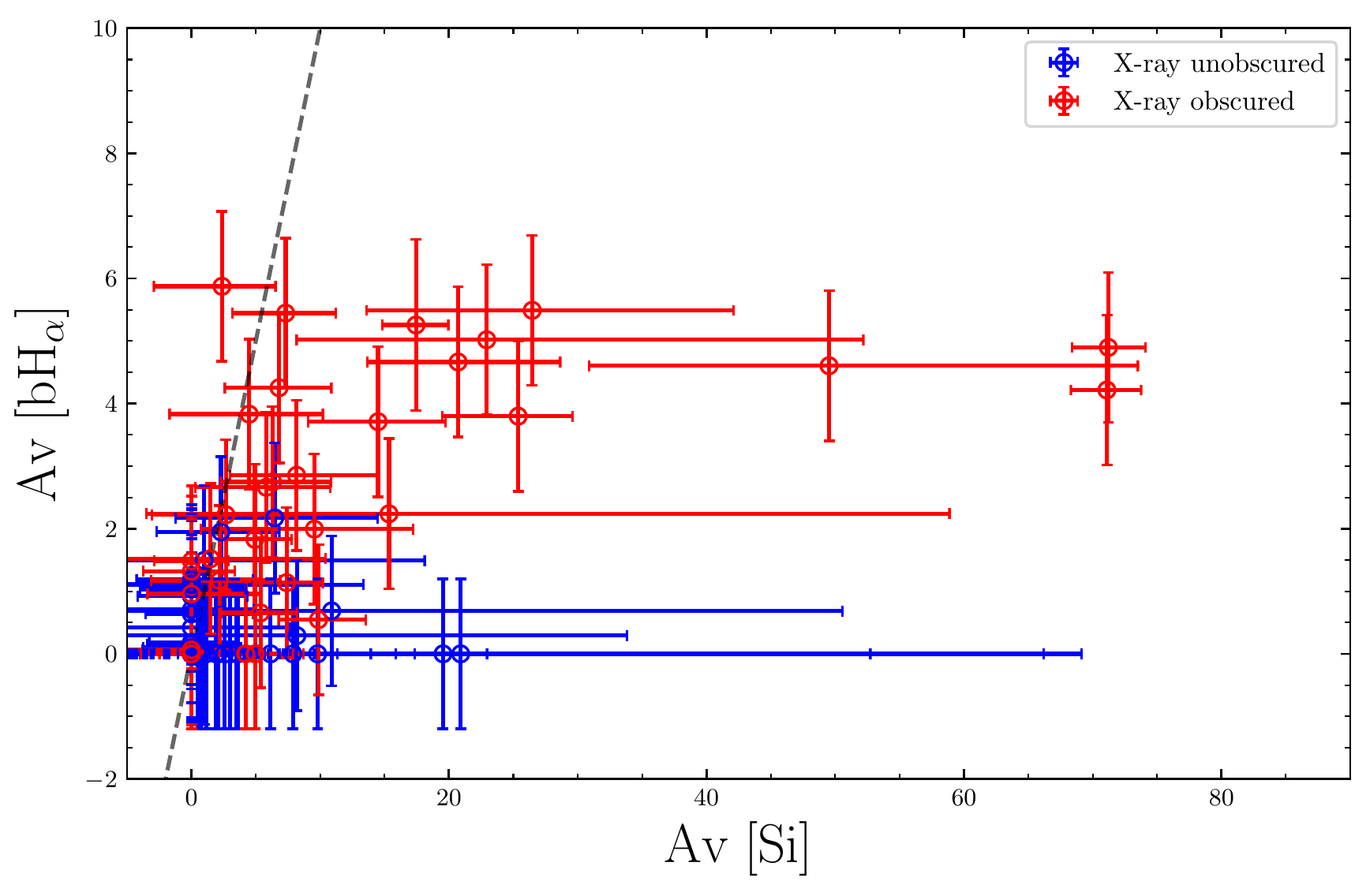}
	\caption{The optical extinction of the broad $H\alpha$ line ($A_{V}[\mathrm{bH}\alpha]$) versus the silicate strength-inferred 
	$V$-band extinction ($A_V[\mathrm{Si}]$). The blue and red symbols indicate X-ray unobscured (i.e., $N_\mathrm{H}^\mathrm{X} 
	<10^{21.5}\ \mathrm{cm^{-2}}$) and obscured sources (i.e., $N_\mathrm{H}^\mathrm{X}\geq 10^{21.5}\ \
	\mathrm{cm^{-2}}$), respectively. Note that the uncertainties of $A_{V}[\mathrm{bH}\alpha]$ are caused by both the measurement 
	errors of the broad $H\alpha$ fluxes and a systematic uncertainty of $1.2$ mag \citep{Shimizu2018}. The dashed line indicates the 
	one-to-one relation. }
	\label{fig:avsil}
\end{figure}

We first investigate the relation between $A_{V}[\mathrm{bH}\alpha]$ and the silicate strength-inferred $V$-band extinction (hereafter 
$A_V[\mathrm{Si}]$) where $A_V[\mathrm{Si}]$ is estimated from $S_{\mathrm{Sil}}$ following the methodology presented in Section 3 
of \cite{Shi2006}. Note that, for sources with silicate emission lines, we set $A_V[\mathrm{Si}]=0$. For the 93 AGNs in our final sample 
that have broad $H\alpha$ measurements, we follow \cite{Merloni2014} and classify them into two categories according to 
$N_\mathrm{H}^\mathrm{X}$, i.e., the X-ray unobscured (i.e., $N_\mathrm{H}^\mathrm{X} < 
10^{21.5}\ \mathrm{cm^{-2}}$) and obscured sources (i.e., $N_\mathrm{H}^\mathrm{X}\geq 10^{21.5}\ \mathrm{cm^{-2}}$). The results 
are shown in Figure~\ref{fig:avsil}. For X-ray unobscured AGNs, their $A_{V}[\mathrm{bH}\alpha]$ and $A_V[\mathrm{Si}]$ values 
are both small. However, for X-ray obscured AGNs, $A_{V}[\mathrm{bH}\alpha]$ and $A_V[\mathrm{Si}]$ are large and the former 
is on average much smaller than the latter (see Section~\ref{sect:add} for the discussions of the possible physical 
reasons). 

\begin{figure}
\includegraphics[width=0.75\textwidth,angle=0]{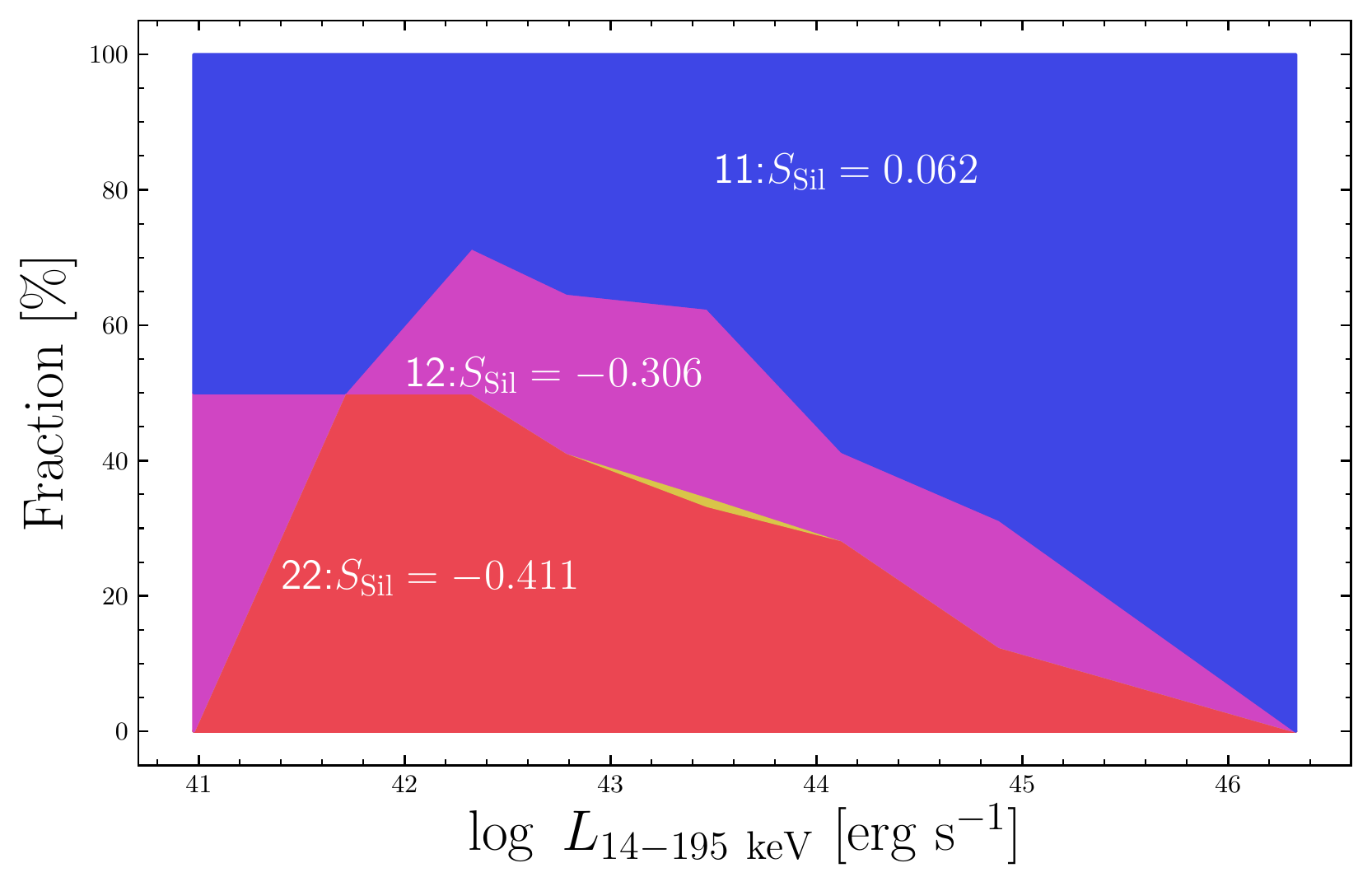}
\caption{Relative fractions of AGN types as a function of the AGN $14$-$195$ keV X-ray luminosity. The blue shaded region represents type-11 
sources (optical type 1 and X-ray unobscured). The purple shaded region represents type-12 sources (optical type 1 and X-ray obscured). The 
yellow shaded region represents type-21 sources (optical type 2 and X-ray unobscured). The red shaded region represents type-22 sources 
(optical type 2 and X-ray obscured). The median silicate strength values for type-11, type-12 and type-22 are annotated, where positive values 
indicate silicate emission and vice versa. There is only one type-21 AGN and its silicate strength is not shown. }
\label{fig:M12}
\end{figure}

Following \cite{Merloni2014}, we further classify our final sample into four categories, i.e., type-11 (optical type 1 and $N_\mathrm{H}^\mathrm{X} 
< 10^{21.5}\ \mathrm{cm^{-2}}$), type-12 (optical type 1 and $N_\mathrm{H}^\mathrm{X}\geq 10^{21.5}\ \mathrm{cm^{-2}}$), type-21 (optical 
type 2 and $N_\mathrm{H}^\mathrm{X}< 10^{21.5}\ \mathrm{cm^{-2}}$), and type-22 (optical type 2 and $N_\mathrm{H}^\mathrm{X}\geq 10^{21.5}\ 
\mathrm{cm^{-2}}$). Their relative fraction as a function of X-ray luminosities are presented in Figure~\ref{fig:M12}. Unlike \cite{Merloni2014}, 
we also can calculate the median value of $S_{\mathrm{Sil}}$ (i.e., $-\tau_{9.7}$) for each type. Consistent with our expectations, type-11/type-22 
sources on average have silicate emission/absorption features. However, type-12 AGNs tend to show prominent silicate absorption, which is 
consistent with the result of Figure~\ref{fig:Shi2006fig}. Among them, the silicate absorption in optical type 1.5 is weak or nearly absent; the silicate 
absorption in optical type 1.9 is rather strong. Our results indicate that, at least for the optical type 1.9 AGNs in the type-12 population, the excess 
of X-ray absorption might not be caused by dust-free broad-line gas. 

In conclusion, the results of Figures~\ref{fig:Shi2006fig}-\ref{fig:M12} along with previous works \citep[e.g.,][]{Shi2006, Merloni2014} suggest that 
the distributions and structures of obscuring gas and extinction dust are very complex.

\section{Discussion}
\label{sect:dis}
As demonstrated in Section~\ref{sect:Results}, $S_{\mathrm{Sil}}$, $N_\mathrm{H}^\mathrm{X}$, and $A_{V}[\mathrm{bH}\alpha]$ often show 
discrepant results, which might be caused by various factors as discussed below. 

\subsection{Smooth torus vs. clumpy torus}
One possible explanation is that the dusty torus is not a smooth ``donut'' but a highly clumpy one. To test this scenario, we compare our 
results in Figures~\ref{fig:Shi2006fig}-\ref{fig:M12} with a smooth torus model of \cite{Fritz2006} and a clumpy torus model of \cite{Nenkova2008p1, 
Nenkova2008p2}, respectively. 
\subsubsection{Testing the smooth torus model}
\label{sect:smooth}
One popular smooth torus model is introduced by \cite{Fritz2006}. In this model, the dust mass density is a function of both radius (with respect to 
the central black hole) and inclination angle ($i$), i.e., 
\begin{equation}
\rho(r, i)=\rho_{0} r^{-q} \mathrm{e}^{-\gamma|\cos (i)|} \\,
\end{equation}
where $\rho_0$ is determined by the equatorial-plane-dust optical depth at $9.7$\mum ($\tau^0_{9.7}$), $q$ is the radial power-law index, and 
$\gamma$ is the polar exponential index, respectively. The torus inner radius is determined by the dust sublimation radius; the ratio of the outer 
radius to the inner radius ($Y$) is allowed to vary as a free parameter. Another parameter is the angular region occupied by the dust ($\Theta$). 
The smooth torus is then illuminated by a central isotropic point AGN emission with a fixed SED of \cite{Schartmann2005}. The radiation emitted 
by the smooth torus is calculated by solving the radiative transfer equations \citep[for more details, see Section~2 of][]{Fritz2006}. 

The explored parameter ranges, which are introduced by \cite{Feltre2012}, are listed in Table~\ref{tbl:smooth}; the covered physical space is 
wider than the original work of \cite{Fritz2006}. For each of the $24000$ smooth-torus SED templates, we first estimate its $S_{\rm Sil}$ by 
following the methodology in Section 5.2 of \cite{Hernan2015}. Second, we calculate the corresponding line-of-sight $N_{\rm H}$ and the 
$V$-band extinction as follows. The line-of-sight optical depth at $9.7$\mum is 
\begin{equation}
\tau_{9.7}= \tau_{9.7}^0 \times\mathrm{e}^{-\gamma|\cos (\theta)|} \\,
\end{equation}
and the corresponding extinction is 
\begin{equation}
\begin{aligned} \frac{A_{9.7}}{\operatorname{mag}} & \equiv 2.5 \log _{10}\left[F_C(\lambda_p) / F(\lambda_p)\right] \\ &=2.5 \log _{10} 
\left[e^{\tau_{9.7}} \right] =1.086 \tau_{9.7} \end{aligned} \\.
\label{eq:A_T}
\end{equation}
The $V$-band extinction is assumed to be $A_{\mathrm{V}}=19\times A_{9.7}$ \citep{Roche1985}. Then, the corresponding line-of-sight 
$N_{\rm H}$ is estimated by considering the dust-to-gas ratio of $A_{\mathrm{V}}/N_{\mathrm{H}}=0.62\times10^{-21}\ \mathrm{mag\;cm^{2}}$ 
\citep{Savage1979}. 

\begin{table}
\bc
\begin{minipage}[]{100mm}
\caption[]{The parameter space of the smooth torus model of \cite{Fritz2006}.\label{tbl:smooth}}\end{minipage}
\setlength{\tabcolsep}{1pt}
\small
  \begin{tabular}{c|c}
   \hline
   \hline
   $\Theta$ (degree) &  $60^\circ$, $100^\circ$, $140^\circ$\\
   \hline
   $q$ &   0.00, 0.25, 0.50, 0.75, 1.0\\
   \hline
   $i$ (degree) & $0^\circ$, $10^\circ$, $20^\circ$ , $30^\circ$, $40^\circ$, $50^\circ$, $60^\circ$, $70^\circ$, $80^\circ$, $90^\circ$\\
   \hline
   $\gamma$ & 0.0, 2.0, 4.0, 6.0\\
   \hline
   $\tau^0_{9.7}$ & 0.1, 0.3, 0.6, 1.0, 2.0, 3.0, 6.0, 10.0\\
   \hline
   $Y$ & 10, 30, 60, 100, 150\\
   \hline
  \end{tabular}
\ec
\tablecomments{0.86\textwidth}{$\Theta$, $q$, $i$, $\gamma$, $\tau^0_{9.7}$, and $Y$ represent the angular region occupied by the dust, 
the radial power-law index, the inclination angle, the polar exponential index, the equatorial-plane-dust optical depth at $9.7$\mum, and the 
ratio of the outer radius to the inner radius, respectively.}
\end{table}

\begin{figure}
\includegraphics[width=0.75\textwidth,angle=0]{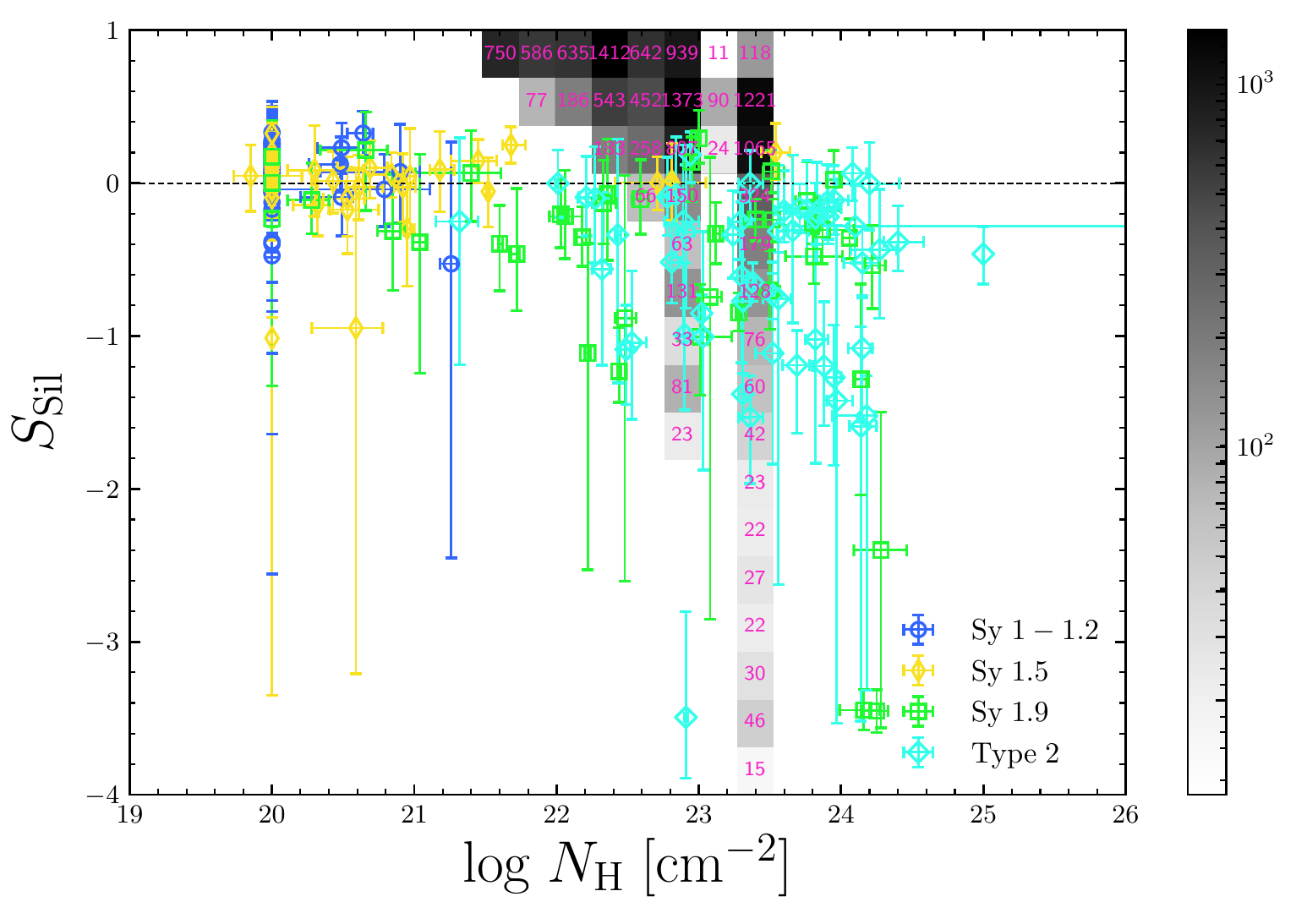}
\caption{$S_{\rm Sil}$ as a function of $N_{\mathrm{H}}$. The blue-open circles, yellow-open-thin diamonds, green-open squares, 
and cyan-open diamonds represent the observations (i.e., $S_{\rm Sil}$ and $N_{\mathrm{H}}^X$) of optical Type 1-1.2, Type 1.5, 
Type 1.9, and Type 2 AGNs, respectively. The grey histogram represents the two-dimensional distribution of $S_{\rm Sil}$ and 
$N_{\mathrm{H}}$ for the smooth torus model of \cite{Fritz2006}. The number of SEDs in each bin is labeled. The smooth torus 
model cannot account for AGNs with small $N_{\mathrm{H}}^{X}$ but evident silicate absorption. }
\label{fig:smooth-nh}
\end{figure}

\begin{figure}
\includegraphics[width=0.75\textwidth,angle=0]{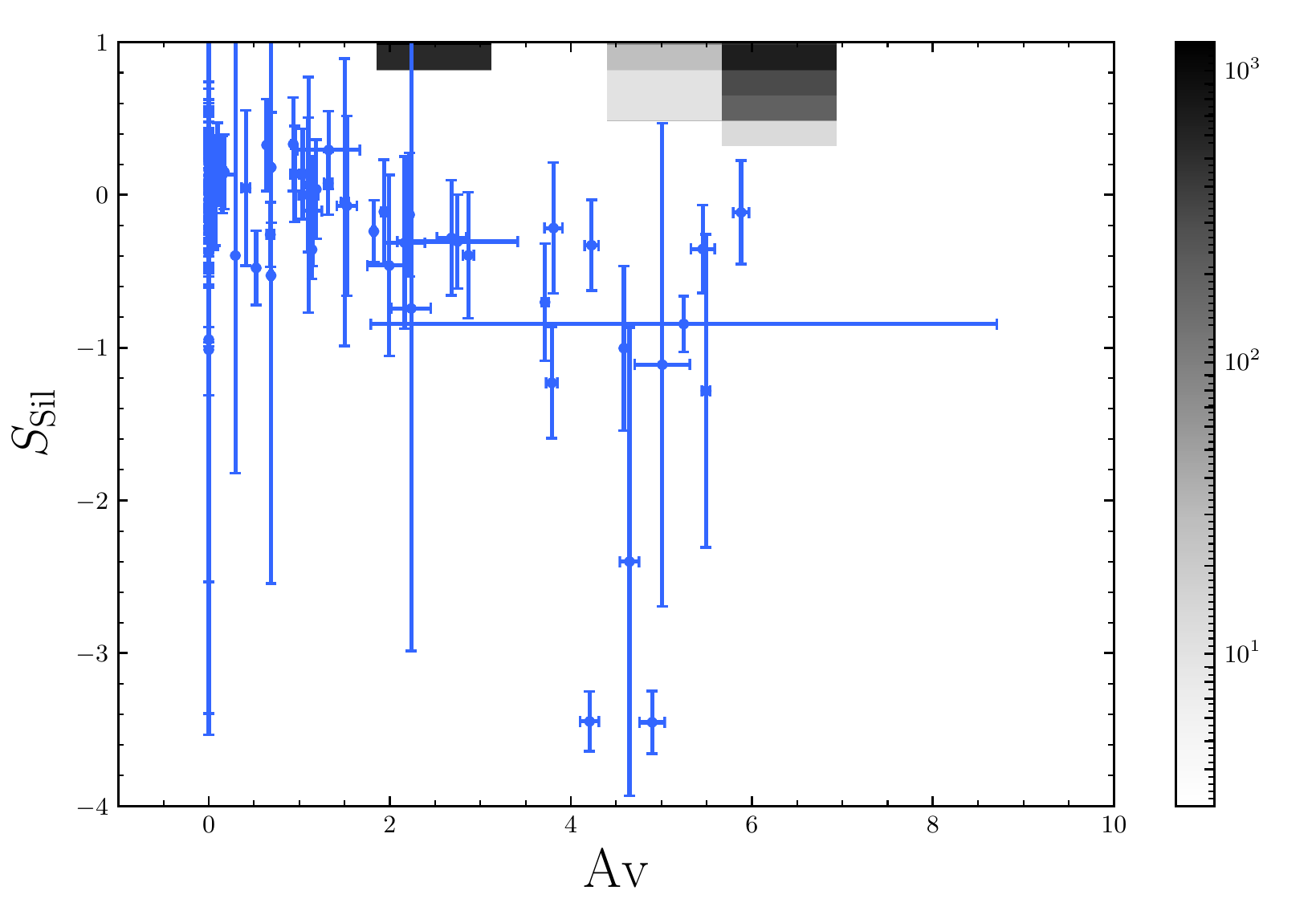}
\caption{$S_{\rm Sil}$ as a function of $A_{\mathrm{V}}$. The blue dots represent the observations (i.e., $S_{\rm Sil}$ and 
$A_{\mathrm{V}}[\rm bH_{\alpha}]$) of our final sample. The grey histogram represents the two-dimensional distribution of 
$S_{\rm Sil}$ and $A_{\mathrm{V}}$ for the smooth torus model of \cite{Fritz2006}. The number of SEDs in each bin is 
indicated by the color bar. The smooth torus model cannot account for AGNs with small $V$-band extinction but evident 
silicate absorption.}
\label{fig:smooth-av}
\end{figure}

The relation between $N_{\mathrm{H}}$ and $S_{\rm Sil}$ for the smooth torus model is presented in Figure~\ref{fig:smooth-nh}. It is 
evident that SEDs with silicate absorption (i.e., negative $S_{\rm Sil}$) always correspond to significant line-of-sight $N_{\mathrm{H}}$ 
(i.e., $>10^{22.5} \mathrm{cm^{-2}}$). Therefore, the smooth torus model cannot explain AGNs with small $N_{\mathrm{H}}^{X}$ but 
evident silicate absorption. The relation between $A_{\mathrm{V}}$ and $S_{\rm Sil}$ is presented in Figure~\ref{fig:smooth-av}. Again, 
the smooth torus model cannot explain our observations. 

\subsubsection{Testing the clumpy torus model}
\label{sect:clumpy}
A clumpy torus model is presented by \cite{Nenkova2008p1, Nenkova2008p2}. According to this model, in the radial direction, dusty 
clouds with the same $V$-band extinction ($\tau_V$) are distributed as $r^{-q}$; in the azimuthal direction, the dusty clouds follow a 
Gaussian distribution, i.e., $e^{-((90-i)/\sigma)^2}$, where $\sigma$ is the Gaussian width. Other parameters are the ratio of the outer 
radius to the inner one ($Y$) and the cloud number along the equatorial plane ($N_0$). The clumpy torus is then illuminated by a 
central AGN source with an SED shape of \cite{Rowan-Robinson1995}. 

\begin{figure}
\includegraphics[width=0.75\textwidth,angle=0]{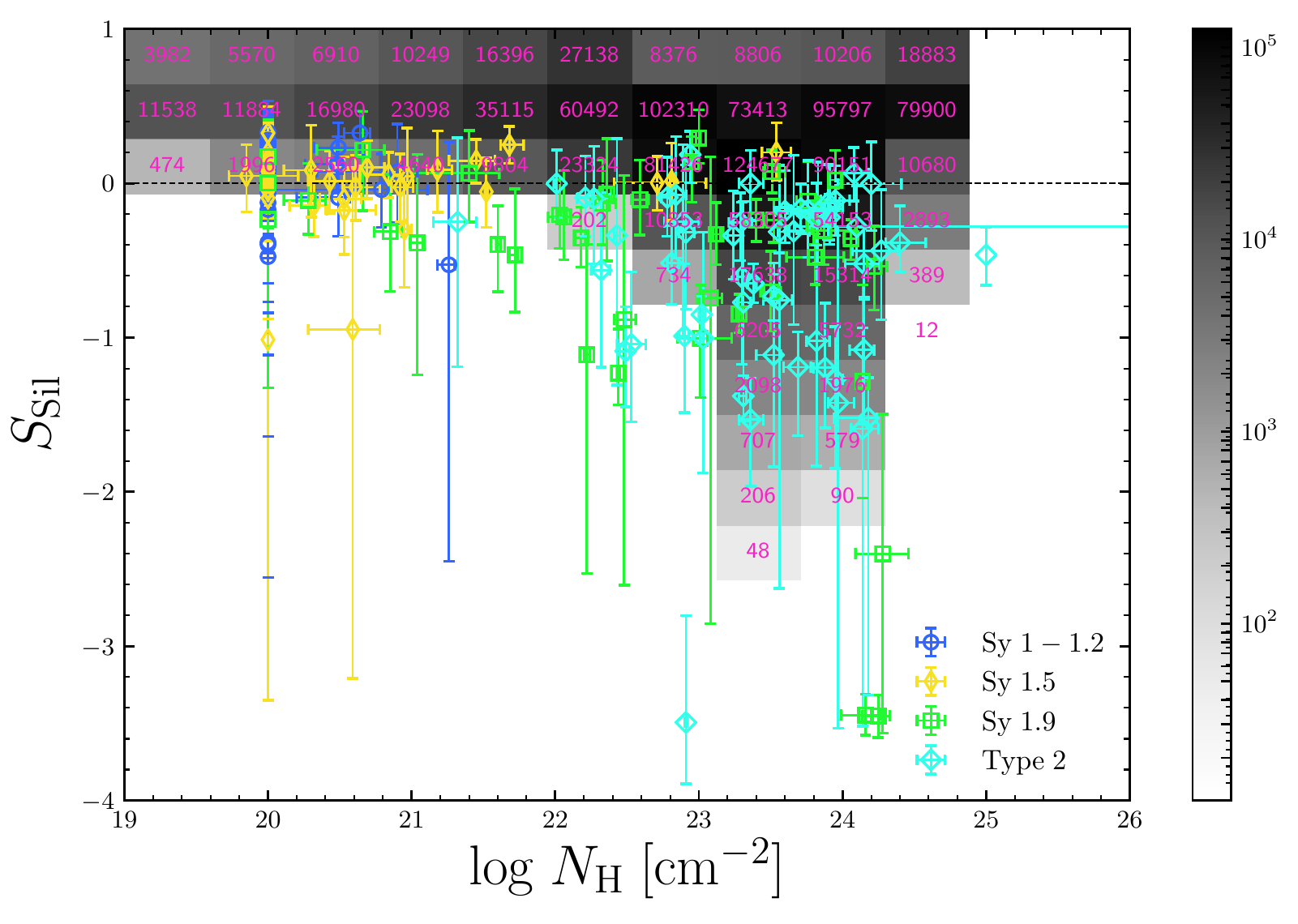}
\caption{Same as Figure~\ref{fig:smooth-nh} but for the clumpy torus model of \cite{Nenkova2008p1}.}
\label{fig:clumpy-nh}
\end{figure}

The explored parameter space of the clumpy torus model is listed in Table~\ref{tbl:clumpy}. Then, for each of the $1247400$ resulting 
SEDs, we estimate the corresponding $S_{\mathrm{Sil}}$. The line-of-sight $N_{\rm H}$ and $V$-band extinction are calculated as 
follows. First, the line-of-sight cloud number is 
\begin{equation}
n_{\mathrm{los}} = N_0 \times \mathrm{e}^{-(\frac{90-i}{\sigma})^2} \\.
\end{equation}
Considering again the dust-to-gas ratio of \cite{Savage1979}, the observed hydrogen column density $N_\mathrm{H}$ is 
\begin{equation}
N_{\mathrm{H}} = \frac{n_{\mathrm{los}}\times 1.086\tau_V}{0.62\times 10^{-21}}\ \mathrm{cm^{-2}} \\.
\end{equation}

\begin{table}
\bc
\begin{minipage}[]{100mm}
\caption[]{The parameter space of the \cite{Nenkova2008p1} clumpy torus template library.\label{tbl:clumpy}}\end{minipage}
\setlength{\tabcolsep}{1pt}
\small
  \begin{tabular}{c|c}
   \hline
   \hline
   $\sigma$ (degree) &  \tabincell{c}{$15^\circ$, $20^\circ$, $25^\circ$, $30^\circ$,$35^\circ$,$40^\circ$,$45^\circ$,$50^\circ$,$55^\circ$,
   \\$60^\circ$,$65^\circ$,$70^\circ$}\\
   \hline
   $q$ &   0.0, 0.5, 1.0, 1.5, 2.0, 2.5, 3.0\\
   \hline
   $i$ (degree) & $0^\circ$, $10^\circ$, $20^\circ$ , $30^\circ$, $40^\circ$, $50^\circ$, $60^\circ$, $70^\circ$, $80^\circ$, $90^\circ$\\
   \hline
   $\tau_\mathrm{v}$ & 10, 20, 40, 60, 80, 120, 160, 200, 300\\
   \hline
   $N_0$ & 1, 2, 3, 4, 5, 6, 7, 8, 9, 
   10, 11, 12, 13, 14, 15\\
   \hline
   $Y$ & 5, 10, 20, 30, 40, 50, 60, 70, 
   80, 90, 100\\
   \hline
  \end{tabular}
\ec
\tablecomments{0.86\textwidth}{$\sigma$, $q$, $i$, $\tau_\mathrm{v}$, $N_0$ and $Y$ represent the 
polar distribution Gaussian-function width, the radial power-law index, the inclination angle, the $V$-band 
optical depth for each dust cloud, the cloud number along the equatorial plane, and the ratio of outer to inner 
radius, respectively. }
\end{table}

\begin{figure}
\includegraphics[width=0.75\textwidth,angle=0]{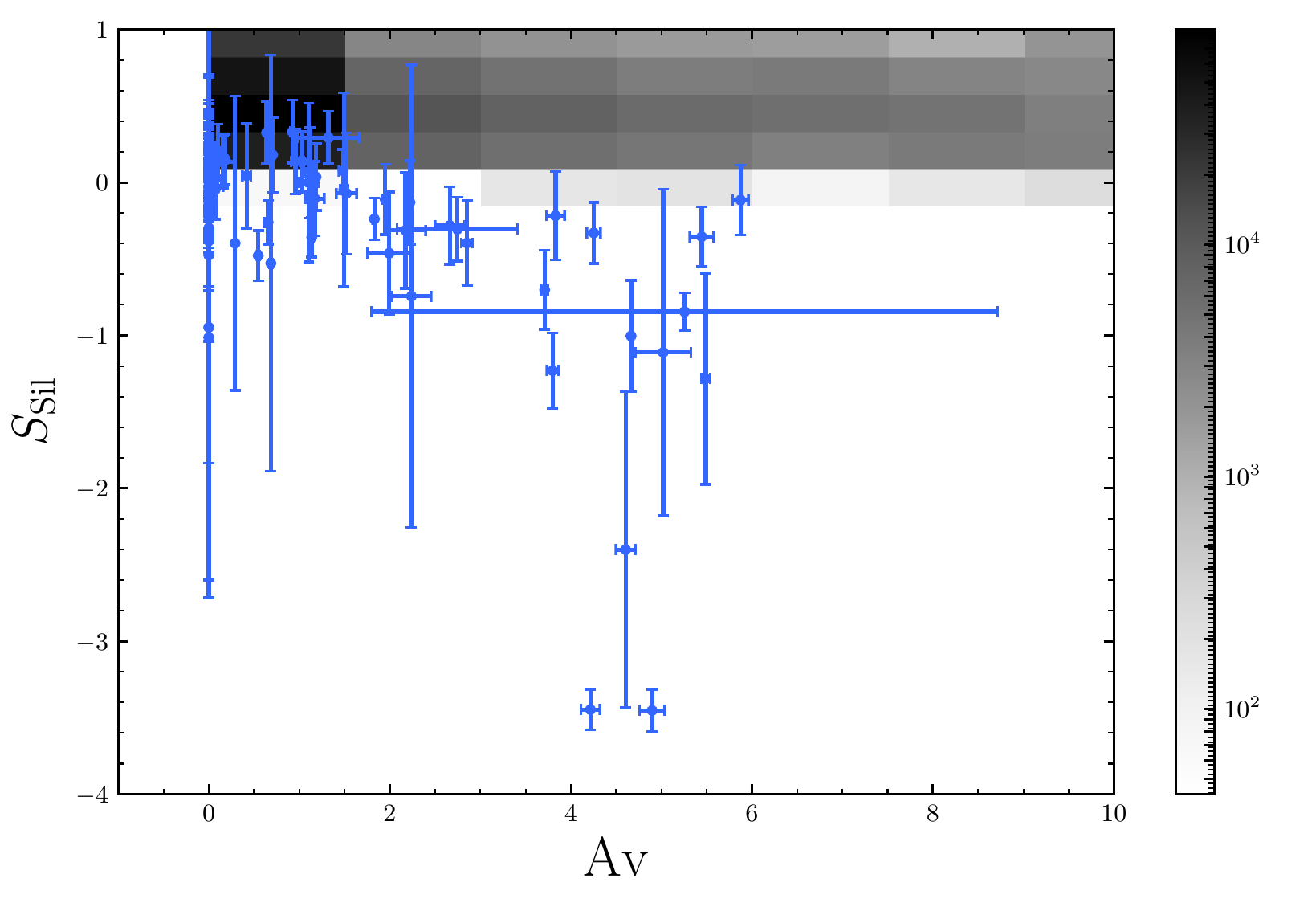}
\caption{Same as Figure~\ref{fig:smooth-av} but for the clumpy torus model of \cite{Nenkova2008p1}.}
\label{fig:clumpy-av}
\end{figure}

The relation between $N_{\mathrm{H}}$ and $S_{\rm Sil}$ ($A_{\mathrm{V}}$ and $S_{\rm Sil}$)  for the clumpy torus model is 
presented in Figure~\ref{fig:clumpy-nh} (Figure~\ref{fig:clumpy-av}). Compared with the smooth torus model, the clumpy torus 
model can explain the observations for most of our sources, especially AGNs with evident silicate absorption and small line-of-sight 
$N_{\mathrm{H}}$ or $V$-band extinction. 

\subsection{Additional gas and dust obscuration}
\label{sect:add}
\begin{figure}
\includegraphics[width=0.75\textwidth,angle=0]{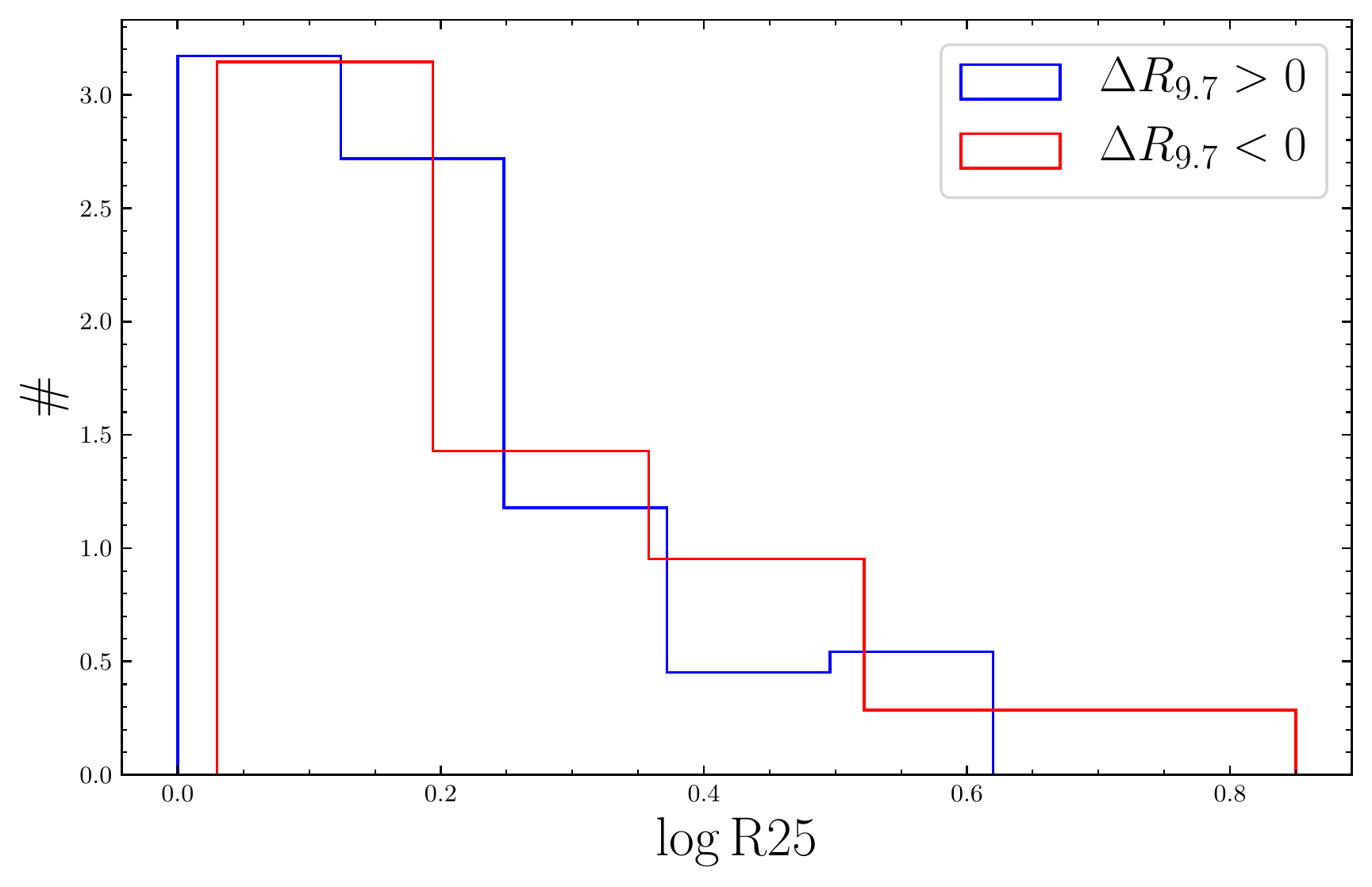}
\caption{Distributions of $\log R25$ for two groups of sources with different $\Delta R_{9.7}$. The distributions are normalized 
to ensure that the total area within each histogram is equal to one. The distribution of $\log R25$ for sources with positive 
$\Delta R_{9.7}$ (i.e., less silicate absorption) is similar to the distribution of $\log R25$ for sources with negative 
$\Delta R_{9.7}$. }
\label{fig:nhx-r25}
\end{figure}

The discrepancy between $S_{\mathrm{Sil}}$ and $N_\mathrm{H}^\mathrm{X}$ or $A_{V}[\mathrm{bH}\alpha]$ might also 
be caused by other effects. For instance, as pointed out by \cite{Goulding2012}, the silicate absorption might be contributed by 
dust located in the host galaxy rather than the AGN torus. If so, we would expect that sources with larger inclination angles should 
have stronger silicate absorption. Following \cite{Goulding2012}, we use the ratio of the major isophotal diameter to the minor one 
(hereafter $R25$) as introduced in the Third Reference Catalog of Bright Galaxies \citep{Corwin1994} to probe the galaxy inclination 
angle (sources with smaller $R25$ values might tend to be more face-on). For the $153$ sources in the final sample, we can obtain 
their $R25$. We then divide these sources into two groups according to their positions relative to the best-fitting relation between 
$R_{9.7}$ and $N_\mathrm{H}^\mathrm{X}$ (see Section~\ref{sect:nhx}). That is, we calculate the difference ($\Delta R_{9.7}$) 
between the observed $R_{9.7}$ and the predicted one from the best-fitting relation. Group 1 (2) sources have positive (negative) 
$\Delta R_{9.7}$. The distributions of $\log R25$ for the two groups are presented in Figure~\ref{fig:nhx-r25}. We also perform the 
Anderson-Darling test to check the differences between the two distributions. We find that the null hypothesis that the two 
distributions are drawn from the same population cannot be rejected (i.e., the $p$-value of the null hypothesis is much larger than 
$0.05$). Therefore, we conclude that the discrepancy between $S_{\mathrm{Sil}}$ and $N_\mathrm{H}^\mathrm{X}$ are unlikely 
to be caused by galaxy-scale dust absorption. 

We then check the discrepancy between $S_{\mathrm{Sil}}$ and $A_{V}[\mathrm{bH}\alpha]$ as a function of $R25$. We 
define $\delta A_{\mathrm{V}}=(A_V[\mathrm{Si}]-A_{V}[\mathrm{bH}\alpha])/A_{V}[\mathrm{bH}\alpha]$ and divide the sources 
with non-zero $A_V[\mathrm{Si}]$ and $A_{V}[\mathrm{bH}\alpha]$ into two groups, i.e., group 1 with $\delta A_{\mathrm{V}}>3$ 
and group 2 with $\delta A_{\mathrm{V}}<3$. Again, we find that the two groups share the same distribution of $\log R25$. Therefore, 
the discrepancy between $S_{\mathrm{Sil}}$ and $A_{V}[\mathrm{bH}\alpha]$ is also unlikely to be mainly driven by 
galaxy-scale dust absorption. 

\begin{figure}
\includegraphics[width=0.75\textwidth,angle=0]{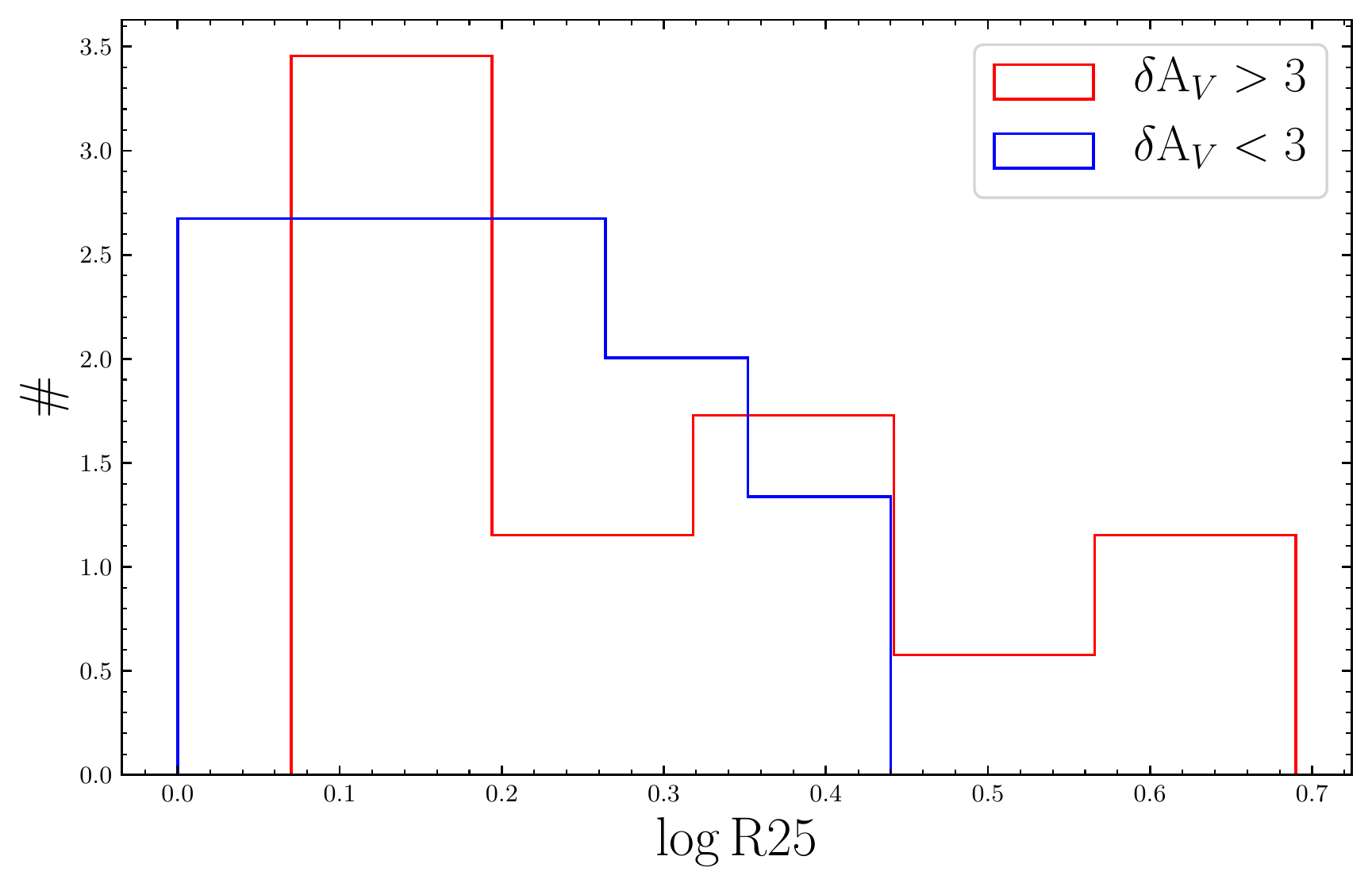}
\caption{Same as Fig.~\ref{fig:nhx-r25}, but for two groups of sources with different $\delta A_{\mathrm{V}}$. The two 
distributions are statistically consistent.} 
\label{fig:ha-r25}
\end{figure}

The discrepancy between $S_{\mathrm{Sil}}$ and $A_{V}[\mathrm{bH}\alpha]$ for X-ray obscured sources might be explained 
as follows. Firstly, as mentioned in Section~\ref{sect:bha}, the estimates of $A_{V}[\mathrm{bH}\alpha]$ rely on the empirical 
relation between X-ray and broad H$\alpha$ luminosities, which is obtained for type 1.0 or 1.2 Seyferts. However, type 1.0 or 1.2 
Seyferts might also suffer dust extinction to some degree; therefore, $A_{V}[\mathrm{bH}\alpha]$ is almost always an under-estimation 
of true dust extinction. Secondly, for X-ray obscured sources, their hidden 
broad emission-line fluxes might be scattered into our line of sight (indeed, spectropolarimetry observations revealed high-polarization 
broad emission lines in at least some of our X-ray obscured sources, e.g., Mrk 3, Mrk 348, and Mrk 1210; see, e.g., 
\citealt{Miller1990, Tran1992}). That is, the observed broad $H\alpha$ flux is larger than the direct broad H$\alpha$ flux (which 
is heavily absorbed by the dusty torus), i.e., $F_{\mathrm{obs}}(\mathrm{bH}\alpha)= 
F_{\mathrm{int}}(\mathrm{bH}\alpha)\exp(-\tau(H\alpha)) + F_{\mathrm{sct}}(\mathrm{bH}\alpha)$, where 
$F_{\mathrm{int}}(\mathrm{bH}\alpha)$, $\tau(H\alpha)$, and $F_{\mathrm{sct}}(\mathrm{bH}\alpha)$ are the intrinsic broad H$\alpha$ 
flux, the optical depth of H$\alpha$, and the scattered broad H$\alpha$ flux, respectively. Note that the scatted light is assumed to 
be not absorbed by the dusty torus. If $\tau$ is larger than $9$ (i.e., $A_V\geq 11.95$ for the extinction law of \citealt{Cardelli1989}), the 
direct broad-line flux is extinguished and $F_{\mathrm{obs}}(\mathrm{bH}\alpha)\cong 
f_{\mathrm{sc}} F_{\mathrm{int}}(\mathrm{bH}\alpha)$, where $f_{\mathrm{sc}}$ is the ratio of $F_{\mathrm{sct}}(\mathrm{bH}\alpha)$ 
to $F_{\mathrm{int}}(\mathrm{bH}\alpha)$. If so, for such sources, the inferred 
$A_{V}[\mathrm{bH}\alpha]$ is about $4.02 + 3.32\log (0.05/f_{\mathrm{sc}})$, i.e., $A_{V}[\mathrm{bH}\alpha]$ might thus be systematically 
underestimated compared to the actual one given that $f_{\mathrm{sc}}$ is about a few percent \citep[e.g.,][]{Reynolds1997}. Thirdly, it is 
also possible that the broad H$\alpha$ in some type 1.9 AGNs might be produced 
by strong outflows (i.e., they are actually type 2 AGNs). One such example is 2MASX J07595347+2323241, whose broad H$\alpha$ is 
surprisingly narrow. In fact, its H$\alpha$-inferred virial black hole mass is two orders of magnitude smaller than the expectation 
of the $M_{\mathrm{BH}}-\sigma$ relation or the infrared broad Pa$\beta$-inferred one \citep{Ricci2017a}. Therefore, possibly due 
to the combined effect of the aforementioned factors, $A_{V}[\mathrm{bH}\alpha]$ is expected to be systematically smaller than the silicate 
strength-inferred $A_{V}$, at least for some of our X-ray obscured AGNs. 

Dust-free gas (e.g., gas in the broad-line regions) can provide additional X-ray obscuration. However, they cannot contribute to 
silicate absorption or $H\alpha$ extinction. As mentioned in Section~\ref{sect:Results} (also see Figures~\ref{fig:Shi2006fig} 
and ~\ref{fig:M12}), dust-free gas cannot fully account for the discrepancy between $S_{\mathrm{Sil}}$ and $N_\mathrm{H}^\mathrm{X}$ 
since many sources are type-2 sources.

\section{Summary of conclusions}
\label{sect:sum}
We decompose the \textit{Spitzer}/IRS spectra of $175$ BASS AGNs to measure their silicate strengths and compare them with 
absorption signatures in X-rays and broad $\rm H\alpha$ emission. Our results are summarized as follows. 
\begin{enumerate}
\item Consistent with previous work \citep{Shi2006}, we confirm a weak relation between the silicate strength and $N_{\mathrm{H}}^X$ 
using more accurate $N_{\rm H}^{\rm X}$ measurements; however, the scatter of the relation is quiet large (see 
Figure~\ref{fig:Shi2006fig} and Section~\ref{sect:nhx}). 

\item For X-ray unobscured AGNs, the silicate strength and the $\rm H\alpha$-inferred $V$-band extinction are both small; 
while for X-ray obscured ones, the silicate strength is much stronger than the expectation of the $\rm H\alpha$-inferred $V$-band 
extinction (see Figure~\ref{fig:avsil} and Section~\ref{sect:bha}). This result and the previous one suggest that the distributions and 
structures of obscuration gas and extinction dust are very complex.

\item We test our data against two popular torus models, i.e., the smooth torus model (see Section~\ref{sect:smooth}) of \cite{Fritz2006} 
and the clumpy torus model (see Section~\ref{sect:clumpy}) of \cite{Nenkova2008p1}. We find that the clumpy torus model is more 
consistent with our observations than the smooth one. 
\end{enumerate}

\section{ACKNOWLEDGEMENTS}
We thank the referee for his/her helpful comments that improved the manuscript.
J.X., M.Y.S., Y.Q.X., and J.Y.L. acknowledge the support from NSFC-11973002, NSFC-11890693, NSFC-11421303, the China 
Postdoctoral Science Foundation (2016M600485), and the CAS Frontier Science Key Research Program (QYZDJ-SSW-SLH006), 
and the K.C. Wong Education Foundation. 

The Combined Atlas of Sources with Spitzer IRS Spectra (CASSIS) is a product of the IRS instrument 
team, supported by NASA and JPL. CASSIS is supported by the "Programme National de Physique Stellaire" 
(PNPS) of CNRS/INSU co-funded by CEA and CNES and through the "Programme National Physique et 
Chimie du Milieu Interstellaire" (PCMI) of CNRS/INSU with INC/INP co-funded by CEA and CNES. 
\label{sect:acknowledgement}

\label{lastpage}


\begin{thebibliography}{99}
  \bibitem[Antonucci(1993)]{Antonucci1993} Antonucci, R. 1993, \araa, 31, 473 

  \bibitem[Baumgartner et al.(2013)]{Baumgartner2013} Baumgartner, W.~H., Tueller, J., Markwardt, C.~B., et al.\ 2013, \apjs, 207, 19 
 
  \bibitem[Burtscher et al.(2016)]{Burtscher2016} Burtscher, L., Davies, R.~I., Graci{\'a}-Carpio, J., et al.\ 2016, \aap, 586, A28
  
  \bibitem[Cardelli et al.(1989)]{Cardelli1989} Cardelli, J.~A., Clayton, G.~C., \& Mathis, J.~S.\ 1989, \apj, 345, 245
  
  \bibitem[Corwin et al.(1994)]{Corwin1994} Corwin, H.~G., Buta, R.~J., \& de Vaucouleurs, G.\ 1994, \aj, 108, 2128
 
 \bibitem[Draine(2011)]{Draine2011} Draine, B.~T.\ 2011, Physics of the Interstellar and Intergalactic Medium by Bruce T. Draine. Princeton University Press
 
  \bibitem[Fritz et al.(2006)]{Fritz2006} Fritz, J., Franceschini, A., \& Hatziminaoglou, E.\ 2006, \mnras, 366, 767 

  \bibitem[Feltre et al.(2012)]{Feltre2012} Feltre, A., Hatziminaoglou, E., Fritz, J., et al.\ 2012, \mnras, 426, 120
  
  \bibitem[Garc{\'\i}a-Burillo et al.(2019)]{Garcia-Burillo2019} Garc{\'\i}a-Burillo, S., Combes, F., Ramos Almeida, C., et al.\ 2019, \aap, 632, A61

  \bibitem[Goulding et al.(2012)]{Goulding2012} Goulding, A.~D., Alexander, D.~M., Bauer, F.~E., et al.\ 2012, \apj, 755, 5

  \bibitem[Gravity Collaboration et al.(2020)]{Gravity2020} Gravity Collaboration, Pfuhl, O., Davies, R., et al.\ 2020, \aap, 634, A1

  \bibitem[Hao et al.(2005)]{Hao2005} Hao, L., Spoon, H.~W.~W., Sloan, G.~C., et al.\ 2005, \apjl, 625, L75

  \bibitem[Hern{\'a}n-Caballero et al.(2015)]{Hernan2015} Hern{\'a}n-Caballero, A., Alonso-Herrero, A., Hatziminaoglou, E., et al.\ 2015, \apj, 803, 109 

  \bibitem[Ichikawa et al.(2019)]{Ichikawa2019} Ichikawa, K., Ricci, C., Ueda, Y., et al.\ 2019, \apj, 870, 31
  
  \bibitem[Imanishi et al.(2018)]{Imanishi2018} Imanishi, M., Nakanishi, K., Izumi, T., et al.\ 2018, \apjl, 853, L25
  
  \bibitem[Jaffarian \& Gaskell(2020)]{Jaffarian2020} Jaffarian, G.~W., \& Gaskell, C.~M.\ 2020, \mnras, 493, 930

  \bibitem[Koss et al.(2017)]{Koss2017} Koss, M., Trakhtenbrot, B., Ricci, C., et al.\ 2017, \apj, 850, 74 

  \bibitem[Lebouteiller et al.(2011)]{Lebouteiller2011} Lebouteiller, V., Barry, D.~J., Spoon, H.~W.~W., et al.\ 2011, \apjs, 196, 8

  \bibitem[Li et al.(2019)]{Li2019} Li, J., Xue, Y., Sun, M., et al.\ 2019, \apj, 877, 5

  \bibitem[Merloni et al.(2014)]{Merloni2014} Merloni, A., Bongiorno, A., Brusa, M., et al.\ 2014, \mnras, 437, 3550
  
  \bibitem[Miller \& Goodrich(1990)]{Miller1990} Miller, J.~S., \& Goodrich, R.~W.\ 1990, \apj, 355, 456

 \bibitem[Mullaney et al.(2011)]{Mullaney2011} Mullaney, J.~R., Alexander, D.~M., Goulding, A.~D., \& Hickox, R.~C.\ 2011, \mnras, 414, 1082 

  \bibitem[Nenkova et al.(2008a)]{Nenkova2008p1} Nenkova, M., Sirocky, M.~M., Ivezi{\'c}, {\v Z}., \& Elitzur, M.\ 2008a, \apj, 685, 147

  \bibitem[Nenkova et al.(2008b)]{Nenkova2008p2} Nenkova, M., Sirocky, M.~M., Nikutta, R., Ivezi{\'c}, {\v Z}., \& Elitzur, M.\ 2008b, \apj, 685, 160 

  \bibitem[Netzer(2015)]{Netzer2015} Netzer, H.\ 2015, \araa, 53, 365

  \bibitem[Nikutta et al.(2009)]{Nikutta2009} Nikutta, R., Elitzur, M., \& Lacy, M.\ 2009, \apj, 707, 1550
  
  \bibitem[Reynolds et al.(1997)]{Reynolds1997} Reynolds, C.~S., Ward, M.~J., Fabian, A.~C., et al.\ 1997, \mnras, 291, 403
  
  \bibitem[Ricci et al.(2015)]{Ricci2015} Ricci, C., Ueda, Y., Koss, M.~J., et al.\ 2015, \apjl, 815, L13

  \bibitem[Ricci et al.(2017)]{Ricci2017a} Ricci, F., La Franca, F., Marconi, A., et al.\ 2017a, \mnras, 471, L41

  \bibitem[Ricci et al.(2017)]{Ricci2017b} Ricci, C., Trakhtenbrot, B., Koss, M.~J., et al.\ 2017b, \apjs, 233, 17
   
  \bibitem[Roche \& Aitken(1985)]{Roche1985} Roche, P.~F., \& Aitken, D.~K.\ 1985, \mnras, 215, 425

  \bibitem[Rowan-Robinson(1995)]{Rowan-Robinson1995} Rowan-Robinson, M.\ 1995, \mnras, 272, 737

  \bibitem[Savage \& Mathis(1979)]{Savage1979} Savage, B.~D., \& Mathis, J.~S.\ 1979, \araa, 17, 73

  \bibitem[Schartmann et al.(2005)]{Schartmann2005} Schartmann, M., Meisenheimer, K., Camenzind, M., et al.\ 2005, \aap, 437, 861

  \bibitem[Schnorr-M{\"u}ller et al.(2016)]{Schnorr2016} Schnorr-M{\"u}ller, A., Davies, R.~I., Korista, K.~T., et al.\ 2016, \mnras, 462, 3570

  \bibitem[Shi et al.(2006)]{Shi2006} Shi, Y., Rieke, G.~H., Hines, D.~C., et al.\ 2006, \apj, 653, 127

  \bibitem[Shimizu et al.(2018)]{Shimizu2018} Shimizu, T.~T., Davies, R.~I., Koss, M., et al.\ 2018, \apj, 856, 154.

  \bibitem[Siebenmorgen et al.(2004)]{Siebenmorgen2004} Siebenmorgen, R., Kr{\"u}gel, E., \& Spoon, H.~W.~W.\ 2004, \aap, 414, 123

  \bibitem[Siebenmorgen et al.(2005)]{Siebenmorgen2005} Siebenmorgen, R., Haas, M., Kr{\"u}gel, E., et al.\ 2005, \aap, 436, L5

  \bibitem[Siebenmorgen et al.(2015)]{Siebenmorgen2015} Siebenmorgen, R., Heymann, F., \& Efstathiou, A.\ 2015, \aap, 583, A120
    
  \bibitem[Sturm et al.(2006)]{Sturm2006} Sturm, E., Hasinger, G., Lehmann, I., et al.\ 2006, \apj, 642, 81
  
  \bibitem[Tran et al.(1992)]{Tran1992} Tran, H.~D., Miller, J.~S., \& Kay, L.~E.\ 1992, \apj, 397, 452

  \bibitem[Urry, \& Padovani(1995)]{Urry1995} Urry, C.~M., \& Padovani, P.\ 1995, \pasp, 107, 803
  
  \bibitem[Yang et al.(2016)]{Yang2016} Yang, G., Brandt, W.~N., Luo, B., et al.\ 2016, \apj, 831, 145
  
\end{thebibliography}
\end{document}